\documentclass[10pt]{iopart}
\usepackage{iopams}
\usepackage{graphicx}
\usepackage{latexsym}

\begin{document}

\title{One step RSB scheme for the rate distortion function}

\author{Tatsuto Murayama and Masato Okada}

\address{RIKEN Brain Science Institute, Wako, Saitama 351-0198, Japan}

\begin{abstract}
We apply statistical mechanics to an inverse problem of linear mapping 
to investigate the physics of the irreversible compression.
We use the replica symmetry breaking (RSB) technique with a toy 
model to demonstrate the Shannon's result. 
The rate distortion function, which is widely known as the theoretical 
limit of the compression with a fidelity criterion, is derived using the
Parisi one step RSB scheme. 
The bound can not be achieved in the sparsely-connected systems,
where suboptimal solutions dominate the capacity.
\end{abstract}




Statistical physics and information
science may have been expected to be directed towards common objectives
since Shannon formulated an information theory based on the concept of 
entropy. 
However, envisaging how this actually happened would have been difficult. 
The situation has greatly been changed since the field of disordered 
statistical systems was maturely established~\cite{book:nishimori}. 
The areas where these relations are particularly strong are 
Shannon's theory~\cite{book:cover-thomas} and the replica theory on classical
spin systems with quenched disorder~\cite{book:dotsenko}. 
Triggered by the work of
Sourlas~\cite{art:sourlas:nature}, these links have recently been 
examined in the area of error corrections~\cite{art:kabashima-saad:replica,
art:kabashima-murayama-saad:codes}, 
network information theory~\cite{art:murayama}, and turbo
decoding~\cite{art:montanari-sourlas}. Recent results of these topics 
are mostly derived using the replica trick. 

However, the research  in the cross-disciplinary field so far can be
categorized as a so-called `zero distortion' decoding scheme in terms of information theory: the system requires perfect reproduction of the input alphabets~\cite{book:cover-thomas}. Here, the same
spin glass techniques should be useful to describe the physics of
systems with a fidelity criterion; i.e., a certain degree of
information distortion is assumed when reproducing the
alphabets. This framework is called the rate distortion 
theory~\cite{art:shannon:distortion,book:berger}. 
Though processing information requires regarding the concept of distortions 
practically, where input alphabets are mostly
represented by continuous variables, statistical physics only employs a few approaches based on highly modified perceptrons~\cite{art:hosaka-kabashima-nishimori}.

In this paper, we introduce a simplified model that achieves the optimality, 
only using parity checks like the Gallager's code~\cite{art:gallager}. 
We, then, can easily see how information distortion can be
handled by the concepts of statistical physics. More specifically, we
study the inverse problem of a Sourlas-type decoding problem by using
the framework of replica symmetry breaking (RSB) of diluted disordered
systems~\cite{book:mezard-parisi-virasoro}. According to our
analysis, this toy model provides an optimal compression scheme for
an arbitrary distortion level, though the encoding procedure
remains an NP-complete problem without any practical encoders at the moment.

The paper is organized as follows. We first review the concept of the rate distortion theory as well
as the main results related to our purpose. We then introduce a toy model. 
Finally we obtain consistent results with information theory. 
Detailed derivations will be reported elsewhere. 

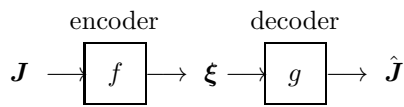
\begin{figure}[h]
\setlength{\unitlength}{1mm}
\begin{center}
\begin{picture}(60,13)(0,12)
  \put(10,13){$\boldsymbol{J} \ \longrightarrow$}
  \put(20,10){\framebox(8,8)}
  \put(23,13){$f$}
  \put(18,20){encoder}
  \put(28,13){$\longrightarrow \ \boldsymbol{\xi} \longrightarrow$}
  \put(44,10){\framebox(8,8)}
  \put(47,13){$g$}
  \put(42,20){decoder}
  \put(52,13){$\longrightarrow \ \hat{\boldsymbol{J}}$}
\end{picture}
\end{center}
\caption{Rate distortion encoder and decoder}
\label{fig:1}
\end{figure}

We start by defining the concepts of the rate distortion
theory and stating the simplest version of the main result. 
Let $J$ be a discrete random variable with alphabet $\mathcal{J}$. 
Assume that we have a source that produces a sequence
$J_1,J_2,\cdots,J_M$, where each symbol is randomly drawn from a
distribution. 
We will assume that the alphabet is finit.
Throughout this paper we use vector notation to represent
sequences for convenience of explanation:
$\boldsymbol{J}=(J_1,J_2,\cdots,J_M)^T \in \mathcal{J}^M$.
Here,  
the encoder describes the source sequence 
$\boldsymbol{J} \in \mathcal{J}^M$ by a codeword 
$\boldsymbol{\xi} = f(\boldsymbol{J}) \in \mathcal{X}^N$. The decoder represents
$\boldsymbol{J}$ by an estimate $\boldsymbol{\hat J}=g(\boldsymbol{\xi})\in
\hat{\mathcal{J}}^M$, as illustrated in Figure~\ref{fig:1}. 
Note that $M$ represents the length of a source
sequence, while $N$ represents the length of a codeword. 
In this case, the rate is defined by $R=N/M$.
Note that the relation $N < M$ always holds when a compression is considered; therefore, $R<1$ also holds. 

A \textit{distortion function} is a mapping
 $d:\mathcal{J}\times \hat{\mathcal{J}} \to \boldsymbol{R}^{+}$ 
 from the set of source alphabet-reproduction alphabet pairs into the set
 of non-negative real numbers.
Intuitively, the distortion $d(J,\hat{J})$ is a measure of the cost of
representing the symbol $J$ by the symbol $\hat{J}$. This definition is
quite general. In most cases, however, the reproduction alphabet
$\hat{\mathcal{J}}$ is the same as the source alphabet
$\mathcal{J}$. Hereafter, we set $\hat{\mathcal{J}}=\mathcal{J}$ and the
following distortion measure is adopted as the fidelity criterion; 
the \textit{Hamming distortion} is given by
 \begin{eqnarray}
  d(J,\hat{J})=\cases{0&for $J = {\hat J}$\\
    1&for $J \neq {\hat J}$\\} \ ,
 \end{eqnarray}
which results in a probable error distortion, since the relation 
$E[d(J,\hat{J})]=P[J \neq \hat{J}]$ holds, where 
$E[\cdot]$ represents the expectation and $P[\cdot]$ the
probability of its argument. The distortion measure is so far defined on
a symbol-by-symbol basis. We extend the definition to sequences. 
The distortion between sequences 
 $\boldsymbol{J}, \boldsymbol{\hat J} \in \mathcal{J}^M$ is  
 defined by 
 $d(\boldsymbol{J},\boldsymbol{\hat J})=(1/M)\sum_{j=1}^M d(J_j,\hat{J}_j)$. 
Therefore, the distortion for a sequence is the average distortion per symbol
of the elements of the sequence. The distortion associated with the code is defined as 
$D=E[d(\boldsymbol{J},\boldsymbol{\hat J})]$, 
where the expectation is with respect to the probability distribution on
 $\mathcal{J}$. 
A rate distortion pair $(R,D)$ should be \textit{achiebable} if
a sequence of rate distortion codes $(f,g)$ exist with
$E[d(\boldsymbol{J},\boldsymbol{\hat J})] \le D$ in the limit $M \to
\infty$. Moreover, the closure of the set of achievable rate distortion
pairs is called the \textit{rate distortion region} for a
source. Finally, we can define a function to describe the boundary;
 the \textit{rate distortion function} $R(D)$ is the infimum of rates
 $R$, so that $(R,D)$ is in the rate distortion region of the source
 for a given distortion $D$.

As in \cite{art:murayama}, we restrict ourselves to a
binary source $J$ with a Hamming distortion measure for simplicity. We assume that
binary alphabets are drawn randomly, i.e., the source is not biased to rule out the possiblity of compression due to redundancy. We
now find the description rate $R(D)$ required to describe the source
with an expected proportion of errors less than or equal to
$D$. In this simplified case, according to Shannon, the boundary can be
written as follows; 
 the rate distortion function for a binary source with Hamming
 distortion is given by
 \begin{eqnarray}
  R(D)=\cases{1-h_2(D)&for $0 \le D \le 1/2$\\
    0&for $1/2 < D$\\} \ ,
 \end{eqnarray}
 where $h_2(\cdot)$ represents the binary entropy function.

Next we introduce a simplified model for the lossy compression. We use
the inverse problem of Sourlas-type decoding to realize the optimal
encoding scheme~\cite{art:sourlas:nature}, a variation of which has recently been investigated by information theorists~\cite{proc:matsunaga-yamamoto}. 
As in the previous paragraphs, we assume that binary alphabets are drawn
randomly from a non-biased source and that the Hamming distortion
measure is selected for the fidelity criterion. 

We take the Boolean representation of the binary alphabet $\mathcal{J}$,
i.e., we set $\mathcal{J}=\{0,1\}$. We also set $\mathcal{X}=\{0,1\}$ to
represent the codewords throughout the rest of this paper. 
Let $\boldsymbol{J}$ be an $M$-bit source sequence,
$\boldsymbol{\xi}$ an $N$-bit codeword, and $\boldsymbol{\hat J}$ an
$M$-bit reproduction sequence. 
Here, the encoding problem can be written as follows.
Given a distortion $D$ and a randomly-constructed Boolean matrix $A$ of
dimensionality $M \times N$, we find the $N$-bit codeword sequence
$\boldsymbol{\xi}$, which satisfies
\begin{eqnarray}
   \boldsymbol{\hat J}&=A\boldsymbol{\xi} \quad \pmod{2} \ , \label{eqn:encoding}
\end{eqnarray}
where the fidelity criterion
$D=E[d(\boldsymbol{J},\boldsymbol{\hat J})]$
holds, according to every $M$-bit source sequence $\boldsymbol{J}$. 
Note that we applied modulo $2$ arithmetics for the additive
operations in (\ref{eqn:encoding}). 
In our framework, decoding will just be a linear mapping 
$\boldsymbol{\hat J}=A\boldsymbol{\xi}$, while encoding remains
an NP-complete problem.

Kabashima and Saad recently expanded on the work of Sourlas, which focused
on the zero-rate limit, to an arbitrary-rate
case~\cite{art:kabashima-saad:replica}.  
We follow their construction of the matrix $A$, so we can treat
non-trivial situations. Let the Boolean
matrix $A$ be characterized by $K$ ones per row and $C$ per column. The finite, and usually small, numbers $K$ and $C$ define a
particular code. The rate of our codes can be set to an arbitrary value
by selecting the combination of $K$ and $C$. We also use $K$ and $C$ as
control parameters to define the rate $R=K/C$. If the value of
$K$ is small, i.e., the relation $K \ll N$ holds, the Boolean matrix $A$
results in a very sparse matrix. 
By contrast, when we consider densely constructed cases, 
$K$ must be extensively big and have a value of $\Or(N)$. 
We can also assume that $K$ is not $\Or(1)$ but $K \ll N$
holds. 

The similarity between codes of this type and Ising spin systems was
first pointed out by Sourlas, who formulated the mapping of a code onto
an Ising spin system Hamiltonian in the context of error
correction~\cite{art:sourlas:nature}. 
To facilitate the current investigation, we first map the problem to that
of an Ising model with finite connectivity following
Sourlasfmethod. We use the Ising representation
$\{1,-1\}$ of the alphabet $\mathcal{J}$ and
$\mathcal{X}$ rather than the Boolean one $\{0,1 \}$; the elements of
the source $\boldsymbol{J}$ and the codeword sequences
$\boldsymbol{\xi}$ are rewritten in Ising values,
and the reproduction
sequence $\boldsymbol{\hat J}$ is generated by taking products of the
relevant binary codeword sequence elements in the Ising representation 
$\hat{J}_\mu=\prod_{i \in \mathcal{S}(\mu)} S_i$.
Here,
we denote the set of codeword indexes $i$ that participate in the message
index $\mu$ by $\mathcal{S}(\mu)=\{i|a_{\mu i}=1\}$ with $A=(a_{\mu i})$.
Therefore, chosen $i$'s correspond to the ones per row, producing a
Ising version of $\boldsymbol{\hat J}$. 
Note that the additive operation in the Boolean representation is
translated into the multiplication in the Ising one.  
Hereafter, we set $J_j,\hat{J}_j,\xi_i = \pm 1$ while we do not change the
notations for simplicity. As we use statistical-mechanics techniques, we
consider the source and codeword-sequence dimensionality ($M$ and $N$,
respectively) to be infinite, keeping the rate $R=N/M$ finite. To
explore the system's capabilities, we examine the Hamiltonian:
\begin{eqnarray}
H(\boldsymbol{S}|\boldsymbol{J})=
\sum_{\mu=1}^M
G[\boldsymbol{S}|J_\mu] \ , \label{H}
\end{eqnarray}
with
\begin{eqnarray}
G[\boldsymbol{S}|J_\mu]
=-J_\mu \prod_{i \in \mathcal{S}(\mu)}S_i
\end{eqnarray}
where we have introduced the dynamical variable $S_i$ to find the optimal
value of $\xi_i$, and $G[\boldsymbol{S}|J_\mu]$ denotes the local connectivity
of a random hypergraph neighboring the message bit $J_\mu$~\cite{art:franz-leone-ricchi-tersenghi-zecchina}.
In addition, we now introduce the connectivity tensor
$\mathcal{G}$ satisfying the relation:
\begin{eqnarray}
\sum_{\mu=1}^M \prod_{i \in \mathcal{S}(\mu)}S_i
=
\sum_{\langle i_1,\cdots,i_K \rangle}
\mathcal{G}_{\langle i_1,\cdots,i_K \rangle}
S_{i_1},\cdots,S_{i_K} \ ,
\end{eqnarray}
for any configuration of $\boldsymbol{S}$.
Elements of the connectivity tensor 
$\mathcal{G}_{\langle i_1,\cdots,i_K \rangle}$ take the value one if the
corresponding indices of codeword bits are chosen (i.e., if all
corresponding indices of the matrix $A$ are one) and zero otherwise; $C$ ones per $i$ index represent the system's degree of
connectivity. 

For calculating the partition function 
$Z(\boldsymbol{J})=\Tr_{\{\boldsymbol{S}\}}\exp[-\beta
\mathcal{H}(\boldsymbol{S}|\boldsymbol{J})]$,
we apply the replica method following the
calculation of Kabashima and Saad~\cite{art:kabashima-saad:replica}. 
To calculate replica free energy, we have to calculate the annealed
average of the $n$-th power of the partition function by preparing $n$
replicas. Here we
introduce the inverse temperature $\beta$, which can be interpreted as a
measure of the system's sensitivity to distortions.
Although larger values of $\beta$ seem to be preferable to
realize smaller reproduction errors, taking the limit $\beta \to \infty$
fails to provide the optimal solution.
This is a direct consequence of the sytem's irreversibility.
As we see in the
following calculation, the optimal value of $\beta$ is naturally
determined when the consistency of the replica symmetry breaking scheme
is considered~\cite{book:mezard-parisi-virasoro,book:dotsenko}. 
We use integral representations of the Dirac $\delta$ function to enforce the
restriction, $C$ bonds per index, on $\mathcal{G}$~\cite{art:wong-sherrington}:
\begin{eqnarray}
 &\delta
  \left(
   \sum_{\langle i_2,i_3,\cdots,i_K \rangle}
   \mathcal{G}_{\langle i,i_2,\cdots,i_K \rangle}
   -C
 \right) \nonumber\\
  =&
  \oint_0^{2\pi}
  \frac{dZ}{2\pi}
  Z^{-(C+1)}
  Z^{\sum_{\langle i_2,i_3,\cdots,i_K \rangle}\mathcal{G}_{\langle i,i_2,\cdots,i_K \rangle}} \ ,
\end{eqnarray}
giving rise to a set of order parameters
\begin{eqnarray}
 q_{\alpha,\beta,\cdots,\gamma}
  =\frac{1}{N}
  \sum_{i=1}^N
  Z_i S_i^{\alpha} S_i^{\beta} \cdots S_i^{\gamma} \ ,
\end{eqnarray}
where $\alpha,\beta,\cdots,\gamma$ represent replica indices, and the
average over $\boldsymbol{J}$ is taken with respect to the probability distribution:
\begin{eqnarray}
 P[J_\mu]
  =\frac{1}{2}\delta(J_\mu-1)
  +\frac{1}{2}\delta(J_\mu+1)
\end{eqnarray}
as we consider the non-biased source sequences for simplicity. 
Assuming the replica symmetry, we use a different representation for the
order parameters and the related conjugate
variables~\cite{art:wong-sherrington}: 
\begin{eqnarray}
 q_{\alpha,\beta,\cdots,\gamma}&=q \int \pi(x)dx \tanh^l (\beta x) \ , \\
 \hat{q}_{\alpha,\beta,\cdots,\gamma}&=\hat{q} \int \hat{\pi}(\hat{x}) d\hat{x}\tanh^l (\beta \hat{x}) \ ,
\end{eqnarray}
where $q=[(K-1)!NC]^{1/K}$ and $\hat{q}=[(K-1)!]^{-1/K}[NC]^{(K-1)/K}$
are normalization constants, and $\pi(x)$ and $\hat{\pi}(\hat{x})$
represent probability distributions related to the integration
variables. 
Here $l$ denotes the number of related replica indices. 
Throughout this paper, integrals with unspecified limits 
denote integrals over the range of $(-\infty,+\infty)$. 
We then obtain an expression for the free energy per source bit
expressed in terms of the probability distributions $\pi(x)$ and $\hat{\pi}(\hat{x})$:
\begin{eqnarray}
  -\beta f
  &=
  \frac{1}{M} \langle \langle \ln Z(\boldsymbol{J}) \rangle \rangle \nonumber\\
  &= \ln \cosh \beta \nonumber\\
  &{\phantom =}+\int \left[ \prod_{l=1}^K \pi(x_l) dx_l \right]
  \left\langle
  \ln
  \left(
  1+\tanh \beta J \prod_{l=1}^K \tanh \beta x_l
  \right)
  \right\rangle_{J} \nonumber\\
  &{\phantom =}-K \int \pi(x) dx \int \hat{\pi}(\hat{x}) d\hat{x} \ 
  \ln(1+\tanh \beta x \ \tanh \beta \hat{x}) \nonumber\\
  &{\phantom =}+\frac{C}{K}\int
  \left[ \prod_{l=1}^C \pi(\hat{x}_l) d\hat{x}_l \right] \ 
  \ln
  \left[
  {\Tr_S}\prod_{l=1}^C (1+S\tanh \beta \hat{x}_l)
  \right] \ , \label{eqn:free}
\end{eqnarray}
where $\langle \langle \cdots \rangle \rangle$ denotes the average over
quenched randomness of $\boldsymbol{J}$, and also $\mathcal{G}$. 
The saddle point equations with respect to probability distributions provide a set
of relations between $\pi(x)$ and $\hat{\pi}(\hat{x})$:
\begin{eqnarray}
  \pi(x)=&\int \left[ \prod_{l=1}^C \pi(\hat{x}_l) d\hat{x}_l \right]
  \delta \left(x-\sum_{l=1}^{C-1} \hat{x}_l \right) \ , \label{eqn:SPE1} \\
  \hat{\pi}(\hat{x})=&\int \left[ \prod_{l=1}^K \pi(x_l) dx_l \right] 
\nonumber\\
  &\times
  \left\langle \delta \left[\hat{x}-\frac{1}{\beta}\tanh^{-1}
  \left(\tanh \beta J \prod_{l=1}^{K-1} \tanh \beta x_l \right)
  \right] \right\rangle_{J} \ .
 \label{eqn:SPE2}
\end{eqnarray}
By using the result obtained for the free energy, we can easily perform
further straightforward calculations to find all the other observable
thermodynamical quantities, including internal energy: 
\begin{eqnarray}
 e
 &=
 \frac{1}{M}
 \left\langle\left\langle
 \Tr_{\boldsymbol{S}}
 H(\boldsymbol{S}|\boldsymbol{J})
 e^{-\beta H(\boldsymbol{S}|\boldsymbol{J})}
 \right\rangle\right\rangle \nonumber\\
 &=
 -\frac{1}{M}
 \frac{\partial}{\partial \beta}
 \langle \langle \ln Z(\boldsymbol{J}) \rangle
 \rangle \ , \label{eqn:energy}
\end{eqnarray}
which records reproduction errors. Therefore, in terms of the
considered replica symmetric ansatz, a complete solution of the problem
seems to be easily obtainable; unfortunately, it is not. 

This set of equations (\ref{eqn:SPE1}) and (\ref{eqn:SPE2}) may be solved numerically for general $\beta$, $K$, and $C$. 
However, there exists an analytical solution of this equations. 
We first consider this case. Two
dominant solutions emerge that correspond to the paramagnetic and the spin glass phases. The paramagnetic solution, which is
also valid for general $\beta$, $K$, and $C$, is in the form of
$\pi(x)=\delta(x)$ and $\hat{\pi}=\delta(\hat{x})$; it has the lowest
possible free energy per bit $f_{\mathrm{PARA}}=-1$, although its entropy
$s_{\mathrm{PARA}}=(R-1)\ln 2$ is positive only for $R \ge 1$. 
It means that the true solution must be somewhere
beyond the replica symmetric ansatz. As a first step, which is
called the one step replica symmetry breaking (RSB), 
$n$ replicas are usually divided into $n/m$ groups, each containing $m$ replicas. 
Pathological aspects due to the replica symmetry may be avoided making
use of the newly-defined freedom $m$.  
Actually, this one step RSB scheme is considered to provide the exact solutions when the random energy model limit is considered~\cite{art:derrida}, 
while our analysis is not restricted to this case so far. 

The spin glass solution can be calculated for both the replica symmetric and
the one step RSB ansatz. The former reduces
to the paramagnetic solution ($f_{\mathrm{RS}}=f_{\mathrm{PARA}}$), which is unphysical
for $R < 1$, while the latter yields $\pi_{\mathrm{1RSB}}(x)=\delta(x)$, 
$\hat{\pi}_{\mathrm{1RSB}}(\hat{x})=\delta(\hat{x})$ with
$m=\beta_g(R)/\beta$ and $\beta_g$ obtained from the root of the
equation enforcing the non-negative replica symmetric entropy 
\begin{eqnarray}
s_{\mathrm{RS}}=\ln \cosh \beta_g-\beta_g \tanh \beta_g+R \ln 2=0 \ , \label{eqn:vanish}
\end{eqnarray}
with a free energy
\begin{eqnarray} 
f_{\mathrm{1RSB}}
=-\frac{1}{\beta_g}\ln \cosh \beta_g-\frac{R}{\beta_g}\ln 2 \ .
\label{eqn:1RSBF}
\end{eqnarray}
The simple expression (\ref{eqn:1RSBF}) is derived analytically without using any approximations.
However, the stability of the solution must be taken into account when considering the validity.

Since the target bit of the estimation in this model is $J_{\langle
i_1,\cdots,i_K \rangle}$ and its estimator the product 
$S_{i_1}\cdots S_{i_K}$, a performance 
measure for the information corruption could be the per-bond energy
$e$. According to the one step RSB framework, the lowest free energy can
be calculated from the probability distributions $\pi_{\mathrm{1RSB}}(x)$ and
$\hat{\pi}_{\mathrm{1RSB}}(\hat{x})$ satisfying the saddle point
equations~(\ref{eqn:SPE1}) and (\ref{eqn:SPE2})
at the characteristic inverse temperature
$\beta_g$, when the replica symmetric entropy $s_{\mathrm{RS}}$
disappears. Therefore, $f_{\mathrm{1RSB}}$ equals $e_{\mathrm{1RSB}}$. 
Let the Hamming distortion be our fidelity criterion. The distortion $D$ 
associated with this code is given by the fraction of the free energies that
arise in the spin glass phase: 
\begin{eqnarray}
 D = \frac{f_{1RSB}-f_{RS}}{2 |f_{RS}|}
 = \frac{1-\tanh \beta_g}{2} \ . \label{eqn:distortion}
\end{eqnarray}
Here, we substitute the spin glass solutions into the expression, 
making use of the fact that the replica symmetric entropy
$s_{\mathrm{RS}}$ disappears at a consistent $\beta_g$, which is  determined by
(\ref{eqn:vanish}).   
Using (\ref{eqn:vanish}) and (\ref{eqn:distortion}), simple algebra
gives the relation between the rate $R=N/M$ and the distortion $D$ in the form
\begin{eqnarray}
R=1-h_2(D) \ ,
\end{eqnarray}
which coincides with the rate distortion function in the 
Shannon's theorem.
We do {\it not} define any non-linear mappings in the decoding stage but 
we implicitly do in the encoding stage. 
This situation is due to the 
duality between channel coding and lossy compression; the channel capacity 
can be achieved using linear encoders.
Furthermore, we
do {\it not} observe any first-order jumps between analytical solutions. 
Recently, we have seen that many approaches to the family of codes, 
characterized by the linear encoding operations, result in a quite 
different picture; the optimal boundary is constructed in the 
random energy model limit and is well captured 
by the concept of a first-order jump. 
Our analysis of this model, viewed as a kind of inverse 
problem, provides an exception. 

\begin{figure}[h]
\begin{center}
\includegraphics[scale=1.0]{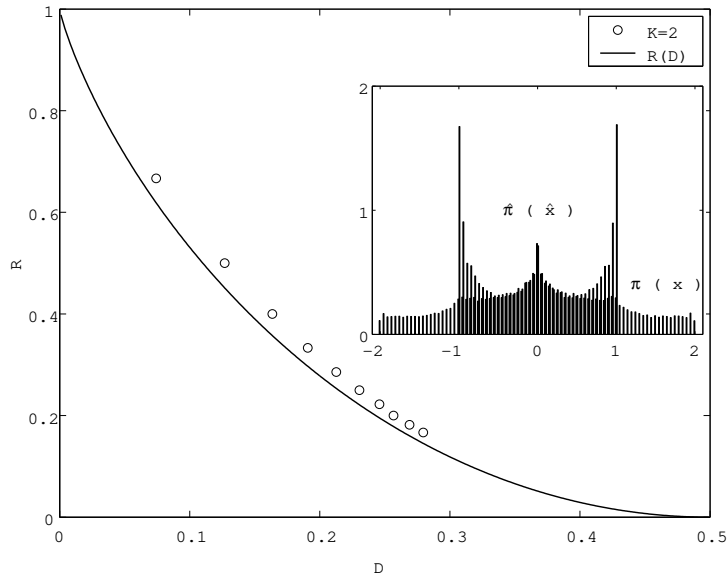}
\end{center}
\caption{Numerically-constructed stable solutions: 
Stable solutions of (\ref{eqn:SPE1}) and (\ref{eqn:SPE2}) for the 
finite values of $K$ and $L$ are calculated by using 
Monte Carlo methods.
We use $10^5$ bin models to approximate the probability 
distributions $\pi_{\mathrm{1RSB}}(x)$ and 
$\hat{\pi}_{\mathrm{1RSB}}(\hat{x})$, 
starting from various initial conditions. 
The distributions converge to the continuous ones, giving suboptimal 
performance.
($\circ$) $K=2$ and $L=3,4,\cdots,12$ ;  
Solid line indicates the rate distortion function $R(D)$.
Inset: Snapshots of the distributions, 
where $L=3$ and $\beta_g=2.35$.
}
\label{fig:2}
\end{figure}

We will now investigate the possiblity of the other solutions satisfying 
(\ref{eqn:SPE1}) and (\ref{eqn:SPE2}) in the case of finite $K$ and $C$. 
Since the saddle point equations appear difficult for 
analytical arguments, we resort to numerical evaluations representing the
probability distributions $\pi_{\mathrm{1RSB}}(x)$ and
$\hat{\pi}_{\mathrm{1RSB}}(\hat{x})$ by up to $10^5$ bin models and
carrying out the integrations by using Monte Carlo methods. Note that the 
characteristic inverse temperature $\beta_g$ is also evaluated
numerically by using (\ref{eqn:free}) and (\ref{eqn:energy}).
We firstly calculate the entropy numerically,
following the basic relation $f=e-\beta^{-1}s$.
Then we choose the proper
value of $\beta$ which provides $s=0$.
We set $K=2$ and selected various 
values of $C$ to demonstrate the performance of stable solutions. 
The numerical results obtained by the one step RSB senario 
show suboptimal properties~[Figure~\ref{fig:2}]. This strongly 
implies that the analytical solution is not the only stable solution.
Furthermore, there has been recent works on the one-step RSB solution
of the model considered in this paper.
The stability of the solution is well examined for some value of $K$ and $C$~\cite{art:montanari-ricci-tersenghi}.

In this paper two points should be noted. 
Firstly, 
we find the consistency between the Shannon's rate distortion theory and 
the Parisi's one step RSB scheme.
Secondly, we confirm that the analytical solution, 
which is consistent with the Shannon's result, 
can not be stable in the sparsely-connected systems.
In case of sparse models,
one might find a polynominal-time algorithm which calculates the suboptimal
solutions, providing a practical method of lossy compression.
We are currently working on the verification.

\ack
The authors thank Yoshiyuki Kabashima and Shun-ichi Amari for their comments 
on the manuscript. 
We also thank Hiroshi Nagaoka and Te Sun Han for giving us valuable references.
This research is, in part, supported by the Special Postdoctoral Researchers Program of RIKEN.

\end{document}